# A Centralized Voltage Controller for Offshore Wind Plants: NY State Grid Case Study

Lin Zhu, Bruno Leonardi, Aboutaleb Haddadi,
Sudipta Dutta, Alberto Del Rosso
Electric Power Research Institute
Knoxville, TN, USA
{lzhu, bleonardi, ahaddadi, sdutta, adelrosso}@epri.com

Victor Paduani, Hossein Hooshyar
New York Power Authority
White Plains, NY, USA
{victor.daldeganpaduani, hossein.hooshyar}@nypa.gov

*Abstract*—This paper proposes a centralized multi-plant reactive power and voltage controller to support voltage control in the interconnected onshore power system. This controller utilizes a hierarchical control structure consisting of a master controller and multiple slave controllers. To validate the proposed method, a realistic planning case of the New York State grid is created for the year 2035, in which nearly 9,500 MW AC or DC connected offshore wind resources are modeled. The performance of the proposed controller is analyzed in the large-scale model under three realistic disturbance scenarios: generator loss, load ramps, and load steps. Results demonstrate how the controller can adequately perform under disturbances to share reactive support proportionally among plants based on their ratings and improve grid voltage stability margins.

## I. INTRODUCTION

Large amounts of offshore wind (OSW) resources are planned to be integrated in North America, especially in the Northeast. According to the Climate Leadership and Community Protection Act (CLCPA) that was passed in 2019, the New York State (NYS) sets ambitious renewable targets by 2035, including the integration of 9,000 MW of OSW [1].

The increased levels of OSW power will have impacts on both the steady-state and dynamic performance of the interconnected onshore power system. One of the main challenges when operating the system under high penetration of OSW generation is to maintain voltage stability and adequate voltage control due to the variability of MW flows near wind power plants. Some of these challenges have been observed in clusters of onshore wind plants. Examples of practical challenges include, for instance, voltage controllers "hunting" or "fighting" each other, reactive power circulation among plants, and plant instability [2]. These challenges are also expected to affect OSW plants, especially if they are electrically close to each other and placed in areas of low short circuit ratio (SCR). Since a large amount of OSW generation is expected to be interconnected into a relatively small area along the New York state coast, the coordination of various voltage control devices becomes important.

There are several voltage control strategies in the literature. For instance, an adaptive droop-based hierarchical optimal voltage control scheme is proposed to coordinate wind turbines (WT) and the WT side voltage-source-converter (VSC) to minimize the voltage deviations of buses inside the WF from the nominal voltage and mitigate reactive power fluctuations of wind turbines in [3]. Furthermore, droop-based control to achieve accurate reactive power sharing in microgrids is proposed in [4]. However, droop-based control may not attain a coordinated performance of multiple actuators, especially in areas of high OSW density. In [5], Guo *et al.* propose a model predictive control-based scheme to coordinate all WTs and WT side VSCs to keep voltages within the feasible range and reduce system power losses. Moreover, a coordinated voltage control scheme for a cluster of offshore wind power plants connected to a VSC-based high-voltage direct current (HVDC) system has been presented in [6]. One critical disadvantage of these methods is that they lack the ability to proportionally share reactive power support by each plant and are more complex to set up if many plants are involved [7],[8]. Consequently, all these methods are tested in use cases including only one OSW, which is insufficient for addressing the previously mentioned challenges.

To address the challenges of coordinating multiple OSW placed in areas of low SCR, this paper proposes a Proportional-Integrator (PI)-based centralized multi-plant reactive power and voltage controller (MPVC) to support voltage coordination in an interconnected onshore power system. It ensures proper reactive power distribution among various actuators or plants. The actuator of the proposed controller is a device with reactive power control capability, such as an onshore/offshore wind plant, a VSC-HVDC terminal, a shunt compensation device, or a synchronous generator. One of the main advantages of the proposed method is a flexible selection of actuators that facilitates the implementation and coordination among different resources (e.g., inverter-based resources and synchronous generators). To validate the effectiveness of the proposed controller for voltage control and voltage stability improvement, in this work, a 1,300 buses transient-stability (TS) model of the New York State grid including 9,000 MW of OSW is developed.

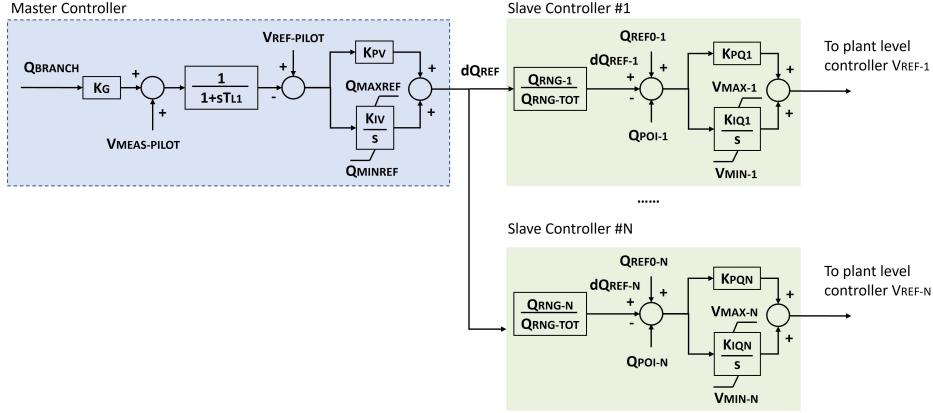

Figure 1. Diagram of the proposed MPVC.

## II. MULTI-PLANT REACTIVE POWER AND VOLTAGE CONTROLLER

The proposed MPVC utilizes a hierarchical control structure which consists of one master controller and multiple slave controllers. The diagram of the proposed MPVC is illustrated in Fig. 1.

The master controller is a PI voltage regulator. It generates the reactive power reference ($Q_{REF}$) based on the difference between the voltage setpoint ($V_{REF-PILOT}$) and the measured voltage ($V_{MEAS-PILOT}$) at the pilot bus. Additionally, it can maintain the reactive power of a selected branch by feedbacking the reactive power of that branch ($Q_{BRANCH}$) as another input signal. The descriptions of the master controller parameters are given in TABLE I.

TABLE I. MPVC MASTER CONTROLLER PARAMETERS

| Variable | Description | Value |
|---|---|---|
| $Q_{BRANCH}$ | Q of monitored branch | N/A |
| $K_G$ | Gain in Q of monitored branch | 0 |
| $V_{MEAS-PILOT}$ | Measured V at pilot bus | N/A |
| $T_{L1}$ | Voltage sensor time constant | 0.02 |
| $V_{REF-PILOT}$ | V setpoint of pilot bus | N/A |
| $K_{PV}$ | Proportional gain in voltage regulator | 4 |
| $K_{IV}$ | Integral gain in voltage regulator | 40 |
| $Q_{MAXREF}$ | Q upper limit (pu, Base = 100 MVA) | 2 |
| $Q_{MINREF}$ | Q lower limit (pu, Base = 100 MVA) | -2 |
| $Q_{REF}$ | Q reference: Master controller output | N/A |

The slave controller is a PI reactive power regulator. The output of the master controller ($dQ_{REF}$) is distributed among multiple slave controllers proportionally to their reactive power ranges at the point of interconnection (POI) buses. Based on the difference between the reactive power reference ($Q_{REF0-i}$ + $dQ_{REF-i}$) and the measured reactive power ($Q_{POI-i}$) injected at the POI bus, the slave controller generates the voltage setpoint ($V_{REF-i}$), which can be added to the voltage setpoint of a renewable plant controller. The descriptions of the slave controller parameters are given in TABLE II. It should be noted that the parameters in Table I and II are specific to this project and would need to be tuned for application in another instance.

Both the master controller and the slave controller are developed in Fortran programming language and then compiled into .dll files for simulations in PSS/e.

TABLE II. MPVC SLAVE CONTROLLER PARAMETERS

| Variable | Description | Value |
|---|---|---|
| $Q_{RNG-i}$ | Q range at POI of OSW i (MVar) | 395.2 (OSW1) 484.3 (OSW2) |
| $Q_{RNG-TOT}$ | Total Q limit at their POIs (MVar) | 879.5 |
| $dQ_{REF-i}$ | Incremental Q reference of OSW i | N/A |
| $Q_{REF0-i}$ | Original Q reference of OSW i | N/A |
| $Q_{POI-i}$ | Measured Q injected to POI bus i | N/A |
| $K_{PQ-i}$ | Proportional gain in Q regulator of OSW i | 0.001 |
| $K_{IQ-i}$ | Integral gain in Q regulator of OSW i | 0.01 |
| $V_{MAX-i}$ | Voltage upper limit of OSW i (pu) | 1.05 |
| $V_{MIN-i}$ | Voltage lower limit of OSW i (pu) | 0.95 |
| $V_{REF-i}$ | Voltage setpoint of OSW i (pu) | N/A |

## III. DEVELOPMENT OF THE NEW YORK STATE GRID MODEL WITH OFFSHORE WIND PLANTS

A realistic planning case of the New York state grid model is created for the year 2035 starting from an existing 2026 planning case. 10 OSW projects are selected from the publicly available NYISO interconnection queue for this work [9]. The total capacity of the selected OSW projects is 9,500 MW, out of which 2,832 MW is in Long Island, and 6,668 MW in New York City. The case studies presented here are carried out under 9,000 MW of instantaneous OSW injection.

### A. OSW Plant Modeling

The OSW generators are modeled as an aggregated wind machine for each project, connected by the collector system and transmission cables to the onshore POI, whose generic topologies are shown in Fig. 2 and Fig. 3. The type 4 wind machine models (WT4E1 and WT4G1) are used to model OSW generators. For DC connected OSW plants, the VSCDCT model is used to model the VSC-HVDC transmission line. Note that for the AC connected OSW plants, a slack bus with a synchronous generator and a transformer are also added to ensure that the power flow solution can converge in PSS/e (Fig. 3).

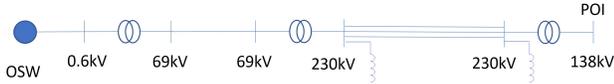

Figure 2. AC connected OSW topology.

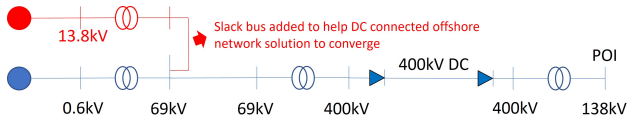

Figure 3. DC connected OSW topology.

The collector system is modeled as one 69 kV cable with aggregated parameters of parallel collectors assuming the length of each collector is 5 miles. The transmission systems are modeled as parallel 230 kV AC cables (for AC connected OSW plants) or one 400 kV DC cable (for DC connected OSW plants). A summary of the cable parameters is provided in TABLE III [10], [11]. The types of connections are determined based on the distances to shore as well as OSW plant ratings. The VSC-HVDC technology is used for cables longer than 60 miles, or when the rating of the OSW plant is higher than 1,000 MW.

The pad mount transformers are modeled with an impedance of 6% on transformer MVA base and an X/R ratio of 8. The substation transformers are modeled with an impedance of 10% on transformer MVA base and an X/R ratio of 50.

TABLE III. CABLE PARAMETERS

| Collector system | | R (ohms/km) | L (mH/km) | C (uF/km) | Current Rating (A) | Voltage Rating (kV) |
|---|---|---|---|---|---|---|
| | | 0.0176 | 0.31 | 0.38 | 820 | 69 |
| Trans. cable | AC | Rac (ohms/km) | Lac (mH/km) | Cac (uF/km) | Current Rating (A) | AC voltage rating (kV) |
| | | 0.0176 | 0.38 | 0.19 | 770 | 230 |
| | DC | Rdc (ohms/km) | Ldc (mH/km) | Cdc (uF/km) | Current Rating (A) | DC voltage rating (kV) |
| | | 0.0132 | 0 | 0 | 1145 | 400 |

IV. SIMULATION RESULTS

The developed MPVC is verified on a 1,300-bus buses New York State grid model including 9,000 MW of OSWs in PSS/e. Note since voltage control is usually a local issue, the original 80,000-bus model is reduced to a 1,300-bus model; however, the New York City and Long Island areas are preserved during the reduction, since that location is where the OSWs are connected to. The power mismatches at boundary buses are converted into equivalent generators that are modeled as classic synchronous generators for dynamic simulation. During development, the model was adjusted to ensure compatibility with a real-time digital simulator, which will be used for controller hardware-in-the-loop tests including NYS grid model [12].

*A. Simulation Setup*

The one-line diagram of the study system is given in Fig. 4. The voltage level of the pilot bus is 138 kV, and two AC-connected OSWs are used as the actuators. Three scenarios are defined to compare the system responses to disturbances with and without the proposed MPVC. In the first scenario, the synchronous generator at a 500 kV bus close to the pilot bus is tripped. In the second scenario, the reactive power of the load at the pilot bus is increased at the rate of 10 MVar/10 s (load was increased by 10 MVar every 10s), and the last scenario is to pick up a 60 Mvar load step directly at the pilot bus. The parameters of the master control and two slave controllers are given in the last column in TABLE I and TABLE II. Please note that the feedback of reactive power of the monitored branch is ignored in this case study by setting $K_G$ to be zero.

Such linear controllers are usually tuned for a range of system operating conditions. The controllers perform well unless system characteristics drastically change. The use of adaptative gains could yield improved performance at the cost of additional design complexity and communication or measurements availability. How to improve controller performance under significant system changes could be a topic of future research for the team.

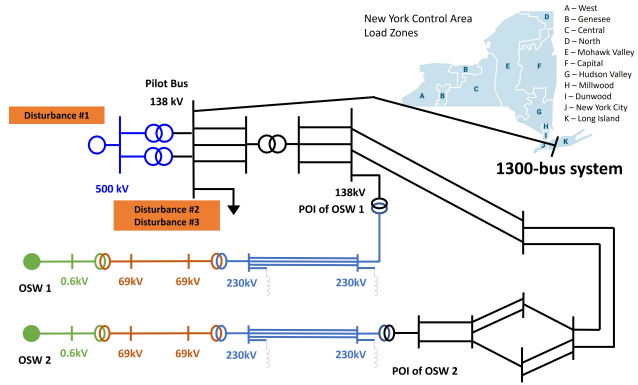

Figure 4. One-line diagram of the study system (1300-bus system).

*B. Test Scenario 1: Loss of Synchronous Generator*

The simulation results of Scenario 1 are given in Figs. 5 to 7. The event is the loss of a nearby synchronous generator at the 500 kV bus at t=5s. Fig. 5 compares the pilot bus voltage with and without the MPVC. It is evident that with the MPVC, the voltage of the pilot bus can be maintained to 1.03 pu which is the original value prior to the disturbance. However, without the MPVC, the voltage drops to 1.028 pu. The output of the master controller is given in Fig. 6. By including the proposed MPVC, the voltage drop caused by the synchronous generator loss is automatically compensated by issuing reactive power requests to the OSWs.

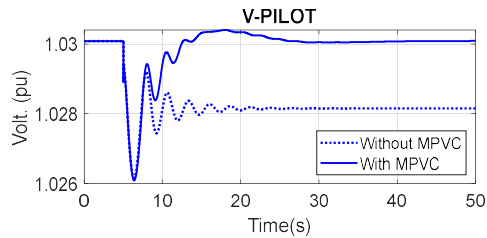

Figure 5. Pilot bus voltage.

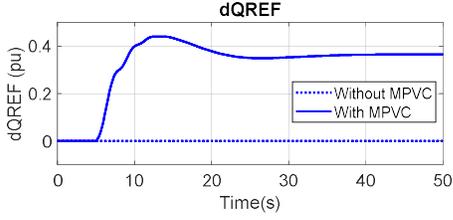

Figure 6. dQ<sub>REF</sub> (output of master controller).

Fig. 7 compares Q<sub>REF</sub> of the slave controller and the actual reactive power injection into the POI bus under two cases: with and without the MPVC. As shown in Fig. 7(a), without the MPVC, The original Q<sub>REF</sub> of OSW 1 is 0.735 p.u. (Base = 100 MVA). Although OSW 1 can inject a certain amount of reactive power to the POI bus by its local Q-V controller, the injected reactive power is not sufficient to bring the pilot bus voltage back to its initial value. On the other hand, when including the proposed MPVC, the injected reactive power to the POI bus by OSW 1 is increased from 0.735 p.u. to 0.896 p.u..

Similarly, Fig. 7(b) shows those variables for OSW 2. With the MPVC, the injected reactive power to the POI bus by OSW 2 is increased from 0.980 p.u. to 1.178 p.u.. Each actuator can successfully follow the control command issued by the associated slave controller. In both cases, the increased reactive power output is proportional to the reactive power range (Q<sub>RNG</sub>) of the two OSWs in TABLE II: 395.2 MVar vs. 484.3 MVar (45% vs. 55%).

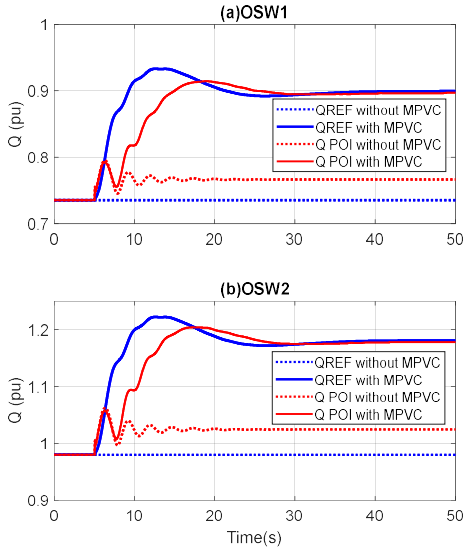

Figure 7. Q<sub>REF</sub> and Q at POI of OSW1 and OSW2.

### C. Test Scenario 2: Load Ramping Up at Pilot Bus

The second simulation scenario is to ramp up the reactive power load at the rate of 10 MVar/10 s at the pilot bus. The upper limit and lower limit of Q<sub>REF</sub> of the master controller are set to be 2 p.u. and -2 p.u., respectively. The simulation results are shown in Figs. 8 to 11. As shown in Fig. 8, when not including the proposed MPVC, the network loses convergence (voltage collapse) at 92.6 s, corresponding to 92.6 MVAr of added load to the pilot bus. Yet, when including the proposed MPVC, the system can supply up to 210.1 MVAr of added load before losing convergence, which occurs at 210.1 s. It is evident that with the MPVC, the system can tolerate a larger reactive power demand at the pilot bus and the system voltage stability is improved. While an event of this nature is not expected in practice, the scenario is useful to demonstrate how the stability margin is increased by the proposed controller. The output of the master controller is given is Fig. 9. The master controller reaches its output upper limit at around 138.6 s and divergence fails at 210.1 s. Fig. 10 shows that each actuator can follow the command given by the slave controller until the network loses convergence.

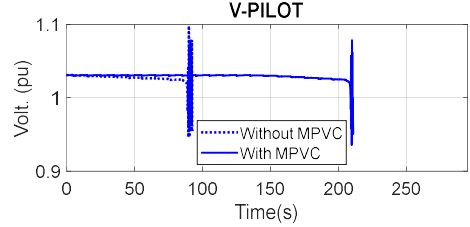

Figure 8. Pilot bus voltage.

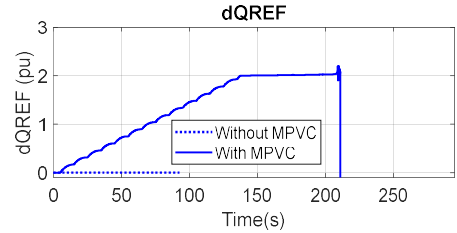

Figure 9. dQ<sub>REF</sub> (output of master controller).

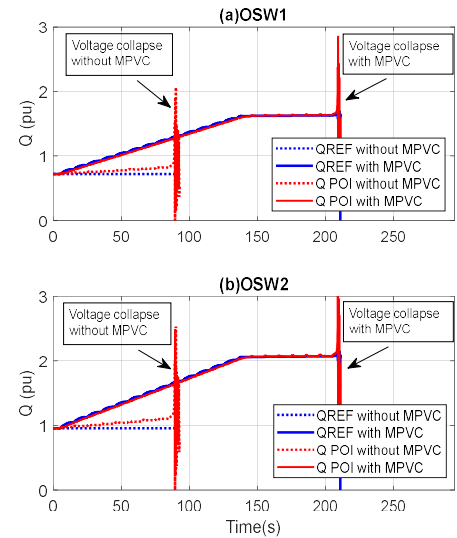

Figure 10. Q<sub>REF</sub> and Q at POI of OSW1 and OSW2.

### D. Test Scenario 3: Direct Load Pickup Up at Pilot Bus

The third scenario is to pick up a 60 MVar reactive power load step at the pilot bus. The simulation results are shown in Figs. 11 to 13. With the MPVC, the pilot bus voltage can return to 1.031 p.u. after the disturbance due to the coordinated control of two actuators. As observed in the previous scenarios, when the MPVC is enabled, each actuator can follow the control commands from the corresponding slave controller. The increased reactive power outputs from each OSW are proportional to their corresponding reactive power ranges in TABLE II: 395.2 MVar vs. 484.3 MVar (45% vs. 55%).

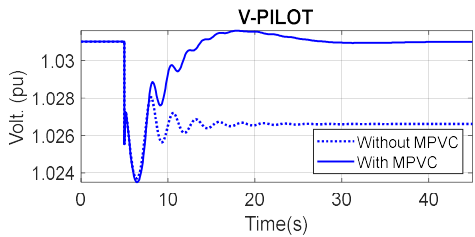

Figure 11. Pilot bus voltage.

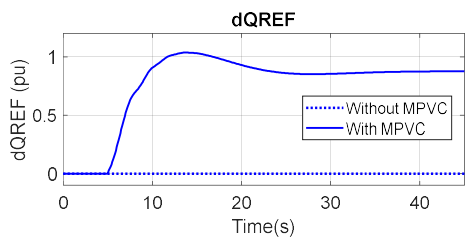

Figure 12. dQ$_{REF}$ (output of master controller).

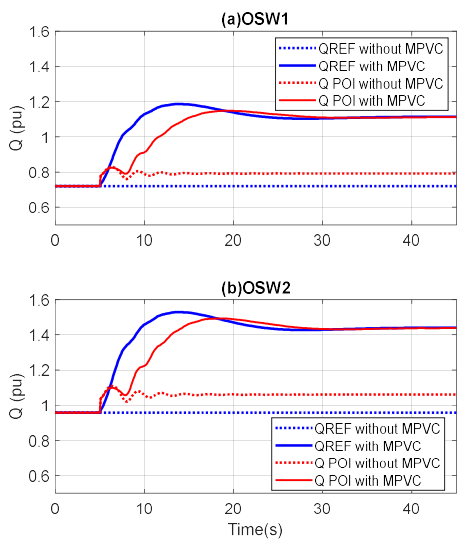

Figure 13. Q$_{REF}$ and Q at POI of OSW1 and OSW 2.

## V. CONCLUSION

This paper presents a case study in which a multi-plant controller that can coordinate multiple OSW plants to support voltage control in the interconnected onshore power system is implemented in a large-scale model of the NYS grid. The controller performance is verified with 9,500 MW (nameplate capacity) AC or DC connected OSW plants that are selected from the publicly available NYISO interconnection queue. Simulation results demonstrate how the centralized controller can assist maintaining the pilot bus voltage at a desired setpoint under grid disturbances. This is achieved by ensuring a proper reactive power distribution among the two actuators based on their reactive power ratings to support voltage and improve voltage stability margins.

As next steps in this work, the authors will run hardware-in-the-loop simulations to study the impacts of communication uncertainties (e.g., time delay, data loss, and loss of actuator) on the proposed controller. The number of actuators will also be increased to further verify the coordination among multiple plants. In addition, the performances the proposed controller and other controllers can be compared.

ACKNOWLEDGMENT

The authors would like to thank Qian Zhang and Tamer Ibrahim for their valuable contributions to this work and comments on this paper.